\documentclass[twocolumn,prd,nofootinbib,aps,floats,floatfix,amsmath,amssymb,secnumarabic]{revtex4} %

\usepackage{graphicx}
\usepackage[usenames,dvipsnames]{color}
\usepackage{amsmath,amssymb}
\usepackage{slashed,verbatim}
\usepackage[colorlinks,citecolor=blue]{hyperref}

\def\sfrac#1#2{{\textstyle{#1\over #2}}}
\newcommand{\be}{\begin{equation}}
\newcommand{\ee}{\end{equation}}
\newcommand{\ba}{\begin{array}}
\newcommand{\ea}{\end{array}}
\newcommand{\bea}{\begin{eqnarray}}
\newcommand{\eea}{\end{eqnarray}}
\newcommand{\sss}{\scriptscriptstyle}
\newcommand{\T}{{\sss T}}
\newcommand{\B}{{\sss B}}
\newcommand{\D}{{\sss D}}
\newcommand{\A}{{\sss A}}

\newcommand{\nc}{\newcommand}
\nc{\ud}{\mathrm{d}}
\nc{\Ego}{E_{\gamma,0}}
\nc{\Eg}{E_{\gamma}}
\nc{\Egp}{E_{\gamma'}}
\nc{\dndE}{\frac{\ud n_e}{\ud E_e}}
\nc{\etal}{\textit{et al.}}
\nc{\sv}{\langle \sigma v \rangle}
\nc{\del}{\partial}
\nc{\nx}{n_{\chi}}
\nc{\mx}{m_{\chi}}
\nc{\z}{{\scriptstyle\cal Z}}
\nc{\bloss}{\frac{4}{3 m_e^2}\sigma_T \rho_{\gamma,0}}

\begin{document}
\title{Cosmological origin of anomalous radio background}
\author{James M.\ Cline}
\email{jcline@physics.mcgill.ca}
\affiliation{Department of Physics, McGill University,
3600 Rue University, Montr\'eal, Qu\'ebec, Canada H3A 2T8}
\author{Aaron C.\ Vincent}
\email{vincent@ific.uv.es}
\affiliation{Instituto de F\'\i sica Corpuscular, Universitat de Val\`encia - CSIC, 
46071, Valencia, Spain}
\begin{abstract}

The ARCADE 2 collaboration has reported a
significant excess in the isotropic radio background, whose
homogeneity cannot be reconciled with clustered sources.  This
suggests a cosmological origin prior to structure formation. We
investigate several potential mechanisms and show that injection of
relativistic electrons through late decays of a metastable particle
can give rise to the observed excess radio spectrum through  
synchrotron emission.  However, constraints from the cosmic microwave
background (CMB) anisotropy, on injection of charged particles and on the
primordial magnetic field, present a challenge. The simplest scenario
is with a $\gtrsim 9$ GeV  particle decaying into $e^+e^-$ at a redshift
of $z\sim 5$, in a magnetic field of $\sim 5 \mu$G, which exceeds the
CMB $B$-field constraints, unless the field was generated after
decoupling. Decays into exotic millicharged particles can alleviate
this tension, if they emit synchroton radiation  in conjunction with a
sufficiently large background magnetic field of a dark U(1)$'$ gauge
field.

\end{abstract}
\maketitle

\section{Introduction}

The ARCADE 2 collaboration has measured the diffuse radio background
at several frequencies between 3 and 90 GHz, with relatively 
small errors up to 10 GHz \cite{Fixsen:2009xn}.  Their measurements
extend earlier ones at lower frequencies $0.01-1$ GHz
\cite{roger}-\cite{reich} that are well fit by a power-law dependence
of the brightness temperature on frequency,
\be
	T \sim T_R(\nu/\nu_0)^\beta
\label{radio_temp}
\ee
with $\beta = -2.6\pm 0.036$. The measured value of $\beta$ is 
within the ballpark of 
expectations ($\beta=-2.7$) for sources in which synchrotron emission
dominates ({\it e.g.,} star-forming galaxies), but the amplitude $T_R
=1.26\pm 0.09$ K (with $\nu_0=1$ GHz)
is $\sim 6$ times  too large to be explained by extrapolations
of populations of known resolved sources \cite{seiffert}.  These
conclusions have received further support in ref.\
\cite{Condon:2012ug}.
Attempts to identify the
excess with previously 
overlooked standard astrophysical sources have so far not met with success
\cite{galactic}-\cite{Vernstrom:2011xt}.
There have been several
attempts to provide a new origin through the annihilation of dark
matter into charged particles, which subsequently undergo synchrotron
emission \cite{Fornengo:2011cn}-\cite{Yang:2012qi}.  It was recently
pointed out that there is a serious challenge to such mechanisms:
the spatial fluctuations in the observed radio signal are much too
small to be consistent with sources that have undergone clustering
\cite{Holder:2012nm}, as would be the case for dark matter
annihilating near the present time.\footnote{Ref.\ \cite{Yang:2012qi} studied
the possibility of dark matter annihilating into charged particles
at earlier times, assuming that ultracompact minihalos and very large
magnetic field exist, finding that  Compton scattering leads
to overproduction of the diffuse x-ray background relative to observed
values.} Ref.\ \cite{Holder:2012nm} conservatively uses the 
linear theory power spectrum for its calculations, whereas
nonlinear structure formation would be expected
to enhance the clustering by a (possibly large) factor that is
currently highly 
uncertain \cite{Ando:2005xg}.  A possible caveat to the analysis of
\cite{Holder:2012nm} is its assumption of Gaussian fields in the computation of error bars,
whereas the intrinsic fluctuations at the relevant scales are likely dominated by
non-linear (hence non-Gaussian) structure.  However, the
measurements are most likely noise-dominated;
thus the statement that clustered
dark matter sources are disfavored by the observed smoothness of
the signal seems to be robust.

These considerations suggest a cosmological origin of the radio
excess, from an epoch prior to the formation of structure and
consequent large  inhomogeneities in the density of decaying dark
matter or of the cosmological magnetic fields. On the other hand, the
excess photons should have been produced relatively late, after
redshift $z=1100$ when recombination occurred, to avoid thermalizing
away such a spectral feature.  When one tries to imagine a mechanism
for producing excess photons with the right spectrum at this epoch or
later, it proves to be highly constrained,  due to cosmic microwave
background (CMB) constraints
$B\lesssim 10^{-9}$G on primordial magnetic fields and on injection of
ionizing charged particles. The purpose of the present work is to
point out some mechanisms that could be promising and the  challenges
that they must be overcome.  We show that a spectrum of excess diffuse
radio background consistent with  observations can be generated either
through  synchrotron emission from
electrons injected into the plasma at some time after decoupling,
depending upon the mass of the decaying particle that produces them. 
However, CMB constraints on the magnetic field (if it was generated
before decoupling) and on the amount of
injected energy turn out to be in conflict with this hypothesis.  

For example, we find that a metastable particle $\chi$ that decays via
$\chi\to e^+e^-$ can produce the desired radio background through
synchrotron emission if  $m_\chi\gtrsim 700$ MeV, consistent with
the CMB constraint on ionizing radiation, but it requires a magnetic
field some 100 times greater than the CMB constraint on primordial
$B$ fields, necessitating a late-time magnetogenesis mechanism.
A more complicated model may be viable: if $\chi$ decays into
millicharged particles that also carry a dark U(1) gauge charge, and if there
is a corresponding background of dark photons with a sufficiently large
dark magnetic field, the CMB constraint on charged particle injection can be
robustly overcome.

Our paper starts with a brief recapitulation of the observed radio
spectrum, reviewing in section \ref{observations} 
the data and the possibility 
that a spectral index
$-2.5$ for the excess temperature provides a consistent description.
In section \ref{motivation} we note that direct production of photons
through decays or annihilations does not give this kind of spectrum,
nor do processes involving
low-energy electrons, 
motivating the alternative of relativistic charged particle injection.
Section \ref{inject} discusses the rate of energy loss of charged
particles through  Compton scattering (CS) and synchrotron emission,
and in it we derive an upper bound on the primordial magnetic field
around the decoupling epoch, from CMB constraints.  This bound
implies that CS is by far the dominant means
of energy loss.  Section \ref{icsect} computes the spectrum of
Compton-scattered photons and consequent bounds from x-ray
observations.
Section
\ref{syncsect} shows that synchrotron emission can produce a radio 
background with the desired properties, if the initial electron
energy is sufficiently large.
In sections
\ref{icsect}-\ref{suddensect} we make a simplifying assumption that 
the charged particles are injected at specific time (sudden decays).
In section \ref{suddensect} we show that this approximation works very
well for decays, and we generalize it to the case of annihilations.
In section \ref{cmbsect} we show that CMB constraints 
are in conflict with the above mechanisms if the injected particles
are electrons, and if the primordial $B$ field was generated before
decoupling.  In section \ref{model} we
suggest a more exotic scenario in which
millicharged particles are instead injected, as a second example of how
the challenges might be overcome.  Conclusions are given in section
\ref{conc}.  A possibly novel way of generating primordial magnetic 
fields (though too weak for our purposes) is elaborated in appendix
\ref{Bdecay}.  In appendix \ref{brem}
we show that brehmsstrahlung cannot yield the observed spectral shape.

\section{The observed spectrum} 
\label{observations}
\begin{figure}[t]
\centering
\includegraphics[width=0.45\textwidth]{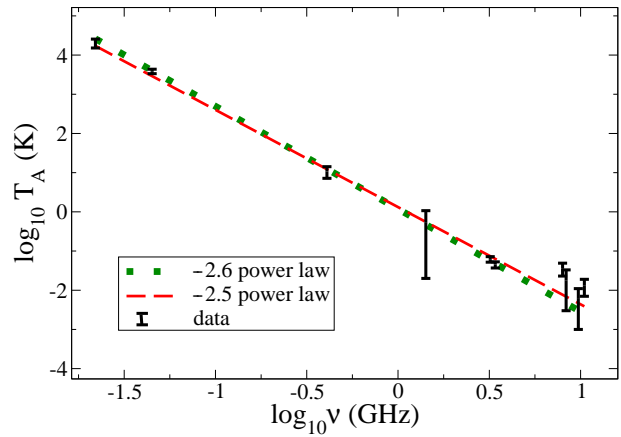}
\caption{Excess antenna temperature $T_\A$ versus frequency as
measured by refs.\  \cite{{Fixsen:2009xn}}- \cite{reich}.  Dotted
and dashed lines show $\nu^{-2.6}$ and $\nu^{-2.5}$ power laws,
respectively.}
\label{data}
\end{figure}

Taking data from table 4 of ref.\ \cite{{Fixsen:2009xn}}, which also
compiles results from the earlier experiments 
\cite{roger}-\cite{reich}, we have replotted the antenna
temperature\footnote{The antenna temperature takes the place of 
$E/(e^{\beta E}-1)$ in the familiar formulas for fluxes, intensities,
{\it etc.,} of blackbody radiation.  We use units $\hbar=c=k_B=1$.} 
$T_\A$ data versus frequency and display them in fig.\ \ref{data}.
The ARCADE-2 data are the 6 highest frequency points (we do not show
those above $10$ GHz, where the error bars are much larger).  Also
shown are $\nu^\beta$ power law dependences, with the ARCADE-2 best
fit value $\beta=-2.6$, and the theoretical prediction $\beta=2.5$
that we will derive in section \ref{syncsect}.  From fig.\
\ref{data}, it is clear that the preference for $\beta=-2.6$ is
driven by the measurements near $\nu = 45$ MHz and 3.3 GHz that have
the smallest errors.  This assumes there is no systematic
miscalibration between the two different experiments  \cite{maeda}
and \cite{{Fixsen:2009xn}} that determined the respective $T_\A$
values.  Even if not, $-2.5$ is within $2.8\sigma$ of 
ARCADE-2's best-fit value.  In the present work, we will
consider $\beta=-2.5$ to be adequate for describing the observations.
It will be seen that there are more challenging problems for finding a
working mechanism than the discrepancy between $-2.5$ and $-2.6$.

\begin{figure}[t]
\centering
\includegraphics[width=0.45\textwidth]{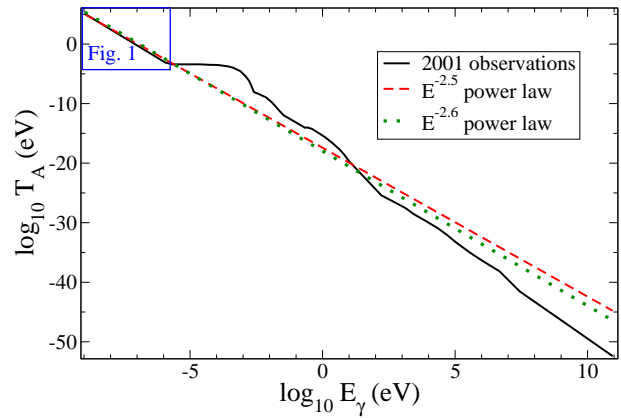}
\caption{Temperature of cosmic background radiation to higher
frequencies/energies, 
compared to $E^{-2.6}$ and $E^{-2.5}$ power laws,  data taken from 
fig.\ 1 of ref.\ \cite{Hauser:2001xs}; lines are our fits to the 
current low-frequency observations.  Region of fig.\ \ref{data} is
shown by box in upper left-hand corner.}
\label{tback}
\end{figure}

To put the radio observations into a broader perspective we have
taken the compendium of diffuse radiation backgrounds presented
in fig.\ 1 of \cite{Hauser:2001xs} and translated it into antenna
temperature using $T_\A = 4\pi^2 \nu^{-2}I_\nu$ to indicate how a new
source in the radio region (energies less than $\sim
10^{-6}$ eV) would compare to higher-energy backgrounds if it
continued with the same power law indefinitely.  At high energies
$E\gtrsim 10$ eV, 
the index $-2.5$ or $-2.6$ would not be sufficiently steep to remain
within observational bounds, but
we will see that this need not be a serious limitation.
First, the mechanism we propose has a natural cutoff energy
above which the spectrum of the new contribution drops sharply, and
this can be dialed by adjusting the initial energy of the injected
charged particles.  Even in the absence of such a cutoff, one
must consider whether the universe is transparent to 
radiation in a given energy range and at a given redshift.
The transparency window as a function of $z$ and $\Eg$ is given in
refs.\ \cite{Chen:2003gz,Slatyer:2009yq}.  At low redshifts, this
window extends from roughly 1 keV to $10^{5}$ GeV, and it is also 
open at $E\lesssim 10$ eV below the energy of Lyman-$\alpha$
absorption.  Thus any x-rays 
produced in conjunction with the radio excess
will be absorbed by the baryonic
plasma for energies below $\sim 1$ keV.  At higher redshifts this
cutoff increases to $\sim 10$ keV.  Depending upon the details,
this effect could weaken the constraining power of experiments
like Chandra \cite{Hickox:2007gj} and XMM-Newton \cite{xmm-newton} that
we will discuss in section \ref{icsect}.

\section{Motivation for relativistic charged particle injection}
\label{motivation}

One might hope to bypass the complications of synchrotron 
and  Compton emission by a mechanism that
directly produces photons after decoupling.  An axion-like particle 
$\chi$ with
a very long lifetime and which decays into two photons is the
simplest possibility.  Although it produces monochromatic photons
at any given time, their energies are redshifted and so a
continuous spectrum results.  Ref.\ \cite{Bertone:2007aw} finds, in 
the case where the decaying particle lifetime 
$\tau_\chi$ is greater than the age of the universe, that the
flux of such photons is given by
\be
	{dJ\over d\Eg} = 2{A\over m_\chi}\left(1 +
\kappa\left(2\Eg\over m_\chi\right)^3\right)^{-\frac12}
\left(2\Eg\over m_\chi\right)^{\frac12}
	\Theta\left(1-{2\Eg\over m_\chi}\right)
\label{decay_spect}
\ee
where $A = 10^{-7}($cm$^2$ s str)$^{-1}(10^{17}$s $/\tau_{\chi})$
(10 GeV/$m_\chi$) assuming $\chi$ is the dark matter, 
and $\kappa = \Omega_\Lambda/\Omega_m \cong 3$.  The energy-dependence can be derived 
from the integral over redshifts \cite{Overduin:2004sz} up to that
at decoupling ($z_{\rm dec}$)
\be
	{dJ\over d\Eg}\sim \int_0^{z_{\rm dec}} dz\, {1\over
H(z)}\,\delta\left[{m_\chi\over
	2\Eg(1+z)} -1\right] 
\label{zint}
\ee
where the Hubble parameter is proportional to $\sqrt{(1+z)^3 +
\kappa}$ and the delta function yields the spectrum of photons from 
a given decay.\footnote{Ref.\ \cite{Overduin:2004sz} derives the spectral
intensity $I(\lambda)$ rather than the flux, 
giving an extra factor of $\lambda^2/\Eg\sim (1+z)^3$ in the
denominator of the integrand.}

\begin{figure}[t]
\centering
\vspace{-0.25cm}
\includegraphics[width=0.45\textwidth]{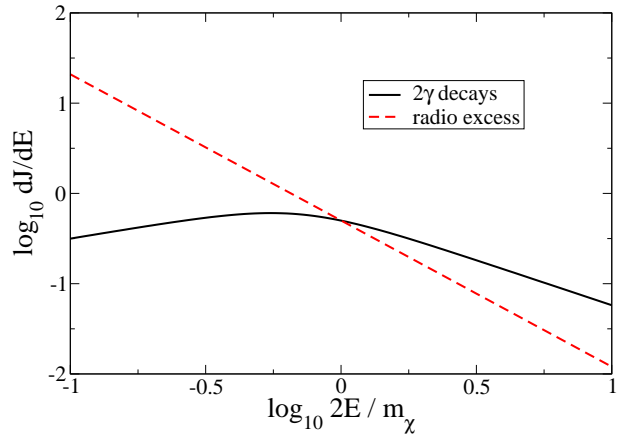}
\caption{Comparison of the shape of the observed radio excess flux (in
arbitrary units) to that produced by dark matter decays $\chi\to 2\gamma$.}
\label{comp}
\end{figure}

On the other hand, the radio
signal of interest has the observed flux
\be
	{dJ\over d\Eg} = {\Eg T(\Eg)\over 2\pi^2} \sim \Eg^{-1.6}
\label{radio_spect}
\ee
where $T(\Eg) \cong T_R\cdot (\Eg/E_0)^{-2.6}$ with
$E_0= 4.1\times 10^{-6}$ eV (corresponding to $\nu=1$ GHz) and $T_R = 1.1\times 10^{-4}$ eV 
from (\ref{radio_temp}).
The shapes of (\ref{decay_spect}) and (\ref{radio_spect}) are compared
in fig.\ \ref{comp}.  The slope of the decay spectrum is never as
negative as the observed one.  We find that the agreement is not
improved by considering decays
into three photons, approximating the spectral distribution as a box
in place of the delta function in (\ref{zint}),
nor by decreasing the lifetime such that the $\chi$'s
disappear before the present time (which also changes the shape of the
present-day spectrum via an extra factor $e^{-(t_0/\tau_\chi)
(1+z)^{-3/2}}$ in (\ref{zint})).\footnote{To avoid the problem of 
large angular
fluctuations in the radio excess, we should insist on lifetimes shorter than the time
scale for structure formation}\ \  Neither do annihilations  
$\chi\chi\to \gamma\gamma$ produce better results, since the input
photon spectrum is monochromatic just like for decays.  
We conclude that particle physics
mechanisms for direct production of photons do not work.  On the other
hand, the radio excess has a spectrum that appears to be consistent
with synchrotron emission.  Production of charged particles thus 
seems like a more promising approach.

In order to generate the correct spectrum, we will furthermore require
the injected charged particles to be relativistic: $E_e \gg m_e$.
The cyclotron output from nonrelativistic electrons in
the cosmic magnetic field will be in the form of peaks around the
gyrofrequency. This is at most $\sim 1$ Hz for values of $B \sim$ nG
allowed by CMB constraints---far below  the MHz-GHz region of the
ARCADE-2 excess.  

Compton scattering with the CMB is a second mechanism which may produce low-energy photons. By
computing the lower-limit to the scattered photon energy, we can show that this mechanism is
not an efficient means of producing radio photons, if the injected electron
is nonrelativistic. A CMB photon with energy $E_{\gamma'} \ll m_e$
scattering head-on with a low energy electron of velocity $v_e \ll 1$ follows the Compton
formula, producing a photon of energy $E_\gamma$:
\begin{equation}
 E_\gamma = \frac{E_{\gamma'}}{1+ \sqrt{\frac{1+v_e}{1-v_e}}\frac{E_{\gamma'}}{m_e}(1-\cos \theta)},
\end{equation}
We are computing an upper limit
to energy loss which we therefore maximise by choosing $\cos \theta = 0$. This yields:
\begin{equation}
 E_\gamma = E_{\gamma'}\left(1 - \sqrt{\frac{1+v_e}{1-v_e}}\frac{E_{\gamma'}}{m_e} +
O\left(\left[\frac{E_{\gamma'}}{m_e}\right]^2\right) \right).
\label{compton}
\end{equation} 
To explain the ARCADE excess, we are seeking a mechanism to produce photons
$E_\gamma \lesssim
0.01 {\overline{E}}_{\gamma'}$ (the average CMB photon energy). This is clearly not possible in the $v_e \ll 1$ regime of
(\ref{compton}). 

\begin{figure*}[t]
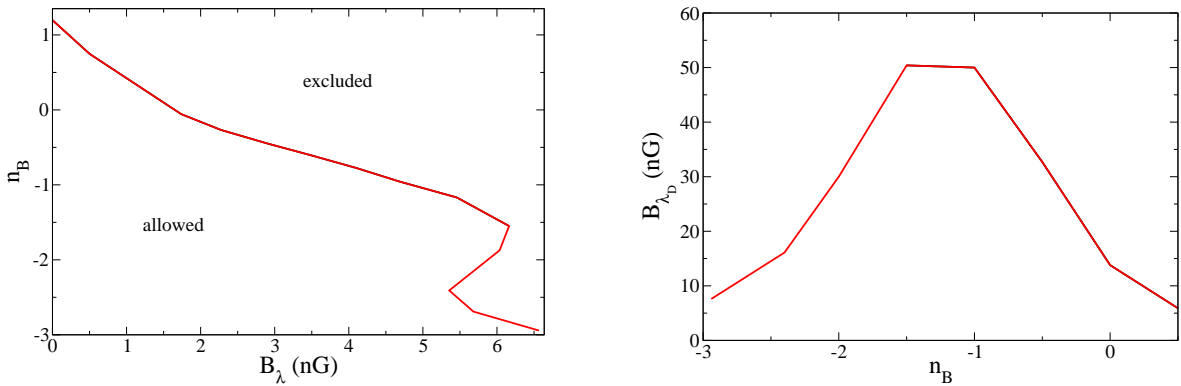

\centering
\vspace{-0.5cm}
\centerline{\includegraphics[width=0.4\textwidth]{finelli-95.eps}
\hfil\includegraphics[width=0.4\textwidth]{Bld.eps}}
\caption{(a) Left: 95\% C.L.\ CMB constraints on $n_B$ versus $B_\lambda$
at $\lambda =1$ Mpc from ref.\ \cite{Paoletti:2010rx}.
(b) Right: $B_{\lambda_D}$ ({\it i.e.,} 
$B_\lambda$ at the damping scale (\ref{damping})) evaluated along
the 95\% C.L.\ contour shown on the left. }
\label{maxB}
\end{figure*}

\section{Injection of electrons and magnetic field constraints}
\label{inject}

We have argued that a more likely mechanism for producing the radio
excess would involve injecting charged particles with excess
energy into the primordial plasma at some time after recombination but
before structure formation, and relying upon the subsequent
synchrotron emission. For example, a metastable particle $\chi$ of
mass $m_\chi> 2 m_e$ could decay into 
$e^+e^-$ pairs at the appropriate time.   Alternatively a first order
phase transition of some scalar field coupled to the Higgs boson could
cause a small change in the Higgs vacuum expectation value, leading to
a change in the mass of the electron and hence its kinetic energy. 
However it seems difficult to obtain relativistic electrons as
required through the latter mechanism.  The  
decaying $\chi\to e^+e^-$ scenario on the other hand naturally
provides relativistic electrons.

It is enlightening to estimate how much energy is needed to produce the observed
excess.  Supposing that the spectral anomaly continues up to some
maximum energy $E_{\rm max}$, its energy density is given by
\bea
	\rho_r &=& {1\over\pi^2}\int_{10^{-2}E_0}^{E_{\rm max}} d\Eg\,
	 \Eg^2\, T(\Eg)
	\nonumber\\ &\cong& 5\times 10^{-21}{\rm eV}^4
	\left({E_{\rm max}\over 10\, E_0}
	\right)^{0.4}
\label{budget}
\eea
(Recall that $E_0$ corresponds to the frequency $\nu_0= 1$ GHz.)
As a fraction of the critical density, this is $\Omega_r =
1.3\times 10^{-10}\, (E_{\rm max}/10\, E_0)^{0.4}$.

Eq.\ (\ref{budget}) gives a lower bound on the excess kinetic energy
in electrons that needs to be injected.  But in fact much more is
required if we rely upon synchrotron emission, since it is not the most efficient means
of dispersing the excess energy in the early universe; 
Compton scattering on CMB photons $(\gamma')$ is much more important.  The total
energy loss rate is due to the sum of the two processes, and their
relative importance depends upon the energy density in CMB photons
versus that in magnetic fields:
\be
	{dE_e\over dt} = \sfrac43\gamma^2 \sigma_\T(\rho_{\gamma'} + \rho_\B)
\label{beq}
\ee
where $\gamma = E_e/m_e$ and 
$\sigma_\T = (8\pi/3)e^4/m_e^2$ is the Thomson cross section.  The relative efficiency for
producing synchrotron radiation is therefore
\be
	{\rho_\B\over \rho_{\gamma'}} \ =\  {\sfrac12 B^2\over
	{\pi^2\over 15} T_{\gamma'}^4}\ \cong\ 10^{-7} \left(B\over
	10^{-9}{\rm\, G} \right)^2
\label{cmb}
\ee
Here we have alluded to the CMB constraint on the comoving magnetic
field $B \lesssim$ $O(10^{-9})$G 
\cite{Durrer:1999bk}-\cite{Paoletti:2010rx}. Taken at face value, a
limit of $B < 1$ nG combined with (\ref{budget}) implies that $10^7$
times more energy in  Compton photons is produced compared to
synchrotron.  If we wish to use the latter to produce the radio 
excess, it would then lead to an extra contribution to $\Omega$ in
photons via CS of order $10^{-3}$, which is 25 times greater than
that in the CMB.  

To understand whether synchrotron emission can be more efficient than
indicated by the above discussion requires us to examine the CMB
contraint on $B$ in greater detail.  The primordial $B$ field is
believed to be stochastic, with a spectrum that in general depends on
the length scale.  The limit $O(10^{-9}){\rm\, G}$ is thus a bound on the
average value $B_\lambda\equiv \langle B(x)B(y)\rangle^{1/2}$ at some
comoving scale $\lambda=|x-y|$, which  is usually taken to be $1$ Mpc
for CMB constraints.   If the spectrum of $B$ is scale invariant
(meaning that $n_\B=-3$ in $B^2_\lambda\sim \lambda^{-(n_\B+3)}$),
then there is no possibility that $B_\lambda$ could be larger
than the nominal bound at some smaller scale.  But if it has a
blue-tilted spectral index $n_\B>-3$, then $B_\lambda$
could conceivably exceed $O(10^{-9}){\rm\, G}$ by some factor at sufficiently small
scales (yet large enough to be coherent over the Larmor radius of an
electron and thus produce synchrotron radiation).

A limitation is that magnetic fields are damped at small
scales by radiative viscosity, and the damping scale is a function of
the $B$ field itself. The result of \cite{Subramanian:1997gi} 
for the damping scale can be expressed as
\cite{Mack:2001gc} 
\be
	{\lambda_\D\over 2\pi} \cong 
	\left({B_\lambda\over 170\times 10^{-9}{\rm\, G}}\right)^{2\over n+5}
	\left({\lambda\over 2\pi\,{\rm Mpc}}\right)^{n+3\over n+5}
	h^{-{1\over n+5}}{\rm\, Mpc}
\label{damping}
\ee
(where $n=n_\B$).
This applies to vector perturbations, which were shown in
ref.\ \cite{Paoletti:2010rx} to dominate
over scalar perturbations.
At $B_{1\rm\, Mpc}=10^{-9}$G, $\lambda_\D$ ranges from $0.05$ to $0.4$ Mpc
as $n_\B$ goes from $-3$ to $2$ (the maximum value usually considered for
a causal mechanism of primordial magnetogenesis).  For a given $n_\B$,
the maximum field value will be attained at the minimum undamped
scale, just above $\lambda_D$.

CMB constraints on $B_\lambda$ and $n_\B$ are correlated, as shown in 
ref.\ \cite{Yamazaki:2010nf,Paoletti:2010rx,Shaw:2010ea}. We reproduce the 
95\% C.L.\ limit on $n_\B$ versus $B_\lambda$ from ref.\ 
\cite{Paoletti:2010rx} in fig.\ \ref{maxB}(a).  The upper limit on
$B_\lambda$ is strongest for the largest values of $n_\B$.
We have evaluated $B_\lambda$ at the damping scale $\lambda_\D$ for
points along this contour to find the maximum allowed $B_\lambda$
as a function of $n_\B$.  The result, plotted in fig.\ \ref{maxB}(b),
is that $B_{\lambda_\D}$ as large as 50 nG is allowed for a spectral
index in the range $n_\B = (-1.5,-1)$, corresponding to damping scales
$\lambda_\D = (0.06-0.1)$ Mpc.  Using this field strength in (\ref{cmb}), the
relative efficiency to produce synchrotron radiation is enhanced by 
a factor of 2500, and the energy density produced by associated 
Compton scattering becomes around 0.01 of that in the CMB, if the
excess radio signal is due to synchrotron emission.  This relaxation
of the nominal bound of $B <$ a few nG is only useful if synchrotron
emission occurs close to the decoupling era since thereafter we expect
the damping scale to increase again as the free charge density
decreases.  This caveat will not affect our conclusions in the end
since we will have to invoke either late-time magnetogenesis or other
exotic particle physics to construct a working scenario.

Stronger limits on $B_\lambda$ for scale-noninvariant spectra can be
obtained by requiring that the energy density $\rho_\B$ in $B$ not exceed the
critical density at earlier times \cite{Caprini:2001nb}; the fraction of energy in 
$\rho_B$ increases at earlier times because $\lambda_D$ depends upon
time (eq.\ (\ref{damping}) is valid around the epoch of
matter-radiation equality).  However these limits would be evaded if
either the time of magnetogenesis was not too early, or if there is an
intrinsic cutoff in the initial spectrum (as one would expect there
must be) that falls below the damping scale before $\rho_\B$ becomes
too large.  Thus the upper limit $B_{\rm max} \cong 50$ nG is fairly
conservative, from the perspective of making the fewest assumptions
about the unknown origin of the primordial magnetic field.  Of course,
if magnetogenesis occurs even later than recombination, we can
evade this constraint as well.  

A single mechanism that could accomplish both of these at once might
interesting 
for explaining the radio excess.  For example, the decay process
$\chi\to e^+e^-$ produces a small electric current at each decay,
which gives rise to stochastic $B$ fields.  We can estimate their
strength as
\be
	\langle B(x)B(y)\rangle\sim {4\pi e^2 n_e\over|x-y|}
\label{decayB}
\ee
for relativistic electrons 
where $n_e$ is the 
electron density. 
In appendix \ref{Bdecay} we derive this, and show that it is
too small for our purposes.
 In the following, we will
assume that magnetogenesis occurred before decoupling, so 
the maximum $B$ field was no more than 50 nG, although the possibility
of relaxing this assumption
should be kept in mind if one is looking for loopholes.

\section{Compton Scattering}
\label{icsect}

Since we have argued that much more energy will be produced by 
Compton scattering than by synchrotron emission, it is necessary to
compute the spectrum of CS photons resulting from injected electrons. 
Although this spectrum itself does not have the right properties to
explain the excess radio background, it has the potential to make a
larger contribution to the diffuse x-ray background than is observed,
and thus provides a constraint on mechanisms that rely upon
synchrotron emission to account for the radio excess.  We derive such
a constraint in this section. 

The rate of Compton emission of photons
of energy $\Eg$ per scattered
electron of initial energy $E_e\gg m_e$, involves an integral over the 
energy spectrum $dn/d\Egp$
of CMB photons:
\be
	{dN_{\gamma}\over dt\,d\Eg} = {2\pi e^4\over
	E_e^2} \int d\Egp 
	{1\over\Egp}{dn\over d\Egp}\, f_{\sss C}(q,\Gamma)\,
	g_{\sss C}(\Eg/\Egp)
\label{icint}
\ee
where (following ref.\ \cite{Jones:1968zza})
\be	
	f_{\sss C}(q,\Gamma) = 2q\ln q + (1+2q)(1-q) + 
	{(1-q)(\Gamma q)^2\over 2(1+ \Gamma q)}
\ee
with $\Gamma = 4 \Egp E_e/m_e^2$ and $q = \Eg/(\Gamma(E_e-\Eg)$
and
\be
	g_{\sss C}(\Eg/\Egp) = \left\{\begin{array}{ll}
	1,& \Eg > \Egp\\
	\Eg/\Egp,& \Eg \le \Egp\end{array}\right.	
\ee
The above parametrization is approximate, and can be inferred from
fig.\ 3 of \cite{Jones:1968zza}; its main purpose is to show that the
spectral index of Compton radiation changes by 1 between the
upscattering and downscattering regimes.
  $f_{\sss C}(q,\Gamma)$, which is meant to describe the regime
where $\Eg > \Egp$, becomes roughly constant at its maximum value
for $\Eg \le \Egp$. This is true provided that the electrons are relativistic. 
The factor $g_{\sss C}(\Eg/\Egp)$ is designed to give the correct
behavior for $\Eg \le \Egp$.
We will further approximate (\ref{icint}) by replacing $E_{\gamma'}$ with its average
value $\overline{E}_{\gamma'} = 2.7\, T_d$, and 
$dn/d\Egp = n_{\gamma,d}/{\overline E}_{\gamma'}$
where $T_d$ is the temperature at
time when the electron was injected (the decay time $t_d$ of $\chi$ in the
model where $\chi\to e^+e^-$) and $n_{\gamma,d}$ is the density of 
CMB photons at this temperature.  Later on we will relax this ``sudden decay
approximation'' by averaging over $t_d$ with the
appropriate dependence of $n_\chi$ on $t_d$.  
The spectrum of emitted photons
at the decay time is then given by
\be
	{dn_{\gamma,d}\over d\Eg} \cong {4\pi e^4 n_{\chi,d}\, n_{\gamma,d}
	\over {\overline E}_{\gamma'}}\, g_{\sss C}(\Eg/{\overline E}_{\gamma'})\int dt\, E_e^{-2}\, f_{\sss C}(q,\Gamma)
\label{dnde}
\ee

\begin{figure}[t]
\centering
\vspace{-0.5cm}
\includegraphics[width=0.45\textwidth]{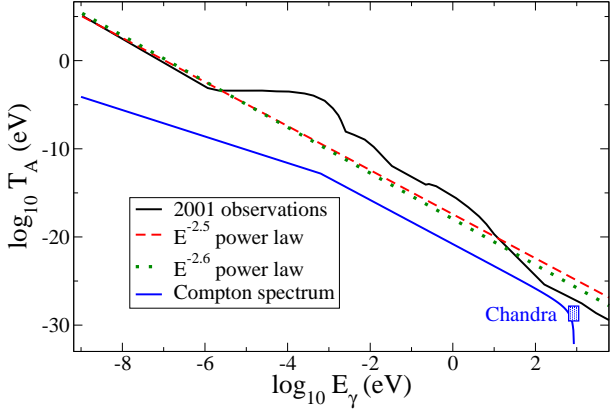}
\caption{Solid (blue): the maximum allowed Compton spectrum consistent
with Chandra constraint, shown for $m_\chi=600$ MeV.  Shaded box shows
lowest-energy Chandra observation \cite{Hickox:2007gj}, near edge of transparency window
below which x-rays are absorbed by intergalactic medium.
Other data taken from fig.\ 1 of ref.\ \cite{Hauser:2001xs}}
\label{ICint}
\end{figure} 

We ignore diffusion since the universe is homogeneous and isotropic
during the epoch of interest, and we assume that the electron spectrum remains
monoenergetic.  
The electron energy diminishes very quickly with time; by solving
eq.\ (\ref{beq}) with $\rho_{\gamma'} \equiv
\rho_{\gamma,d} \gtrsim \rho_\B$, we find
\be
	E_e(t) = {E_d\over 1 +C_e(t-t_d)}
\label{Eeeq}
\ee
for $t>t_d$, where $E_d = m_\chi/2$ is the initial electron energy when it was
injected at time $t_d$, and
\be
	C_e = \frac43 {\sigma_\T\, \rho_{\gamma,d}E_d\over m_e^2}
\ee
The time scale for energy loss is much smaller than the Hubble time,
so we can ignore redshifting of the electron energy.  It is convenient
to change variable $t\to \hat E_e = E_e/E_0$ in the integral (\ref{dnde}), using 
$dE_e/dt = -C_e E_e^2/E_d$, where the minimum electron energy is
$E_0 \equiv \frac12 \Eg[1 + (1 + m_e^2/\Eg{\overline
E}_{\gamma'})^{1/2}]$ from the
kinematics of Compton scattering \cite{Blumenthal:1970gc}.  To evaluate 
${dn_{\gamma}/d\Eg}$ at the present time, we must correct for
redshift by replacing $\Eg \to \z_d \Eg$ and
$\Egp \to \z_d \Egp$ (where the redshift
corresponding to $t_d$ is $z_d\equiv \z_d-1$) in $f_{\sss C}$ and
in $E_0$,
and dividing by an overall factor of $\z_d^2$ (by dimensionality of
${dn_{\gamma}/d\Eg}$).  Thus
\bea
	{dn_{\gamma,0}\over d\Eg} &=& {9\,n_{\chi,0}\, m_e^4\,
	g_{\sss C}
	\over 8\z_d\,(2.7\, T_0)^2\, E_0^3(\z_d)}
	\int_{1}^{E_d\over E_0}
	{d \hat E_e\over \hat E_e^4} f_{\sss C}(q(\z_d),\Gamma)
	\nonumber\\
	&\cong& {9\,n_{\chi,0}\, m_e\, g_{\sss
C}(\Eg/\overline{E}_{\gamma'})
	\over \z_d\,(2.7\, T_0)^{1/2}\, \Eg^{3/2}}\times 0.088
\label{ICres}
\eea
The approximation $0.088$ for the integral holds as long as $E_d/E_0$
is sufficiently larger than 1, {\it i.e.,} for 
\be
\Eg \lesssim 2.6\, T_0(m_\chi/m_e)^2
\label{massbound}
\ee
where $T_0=2.3\times 10^{-4}$ eV is the current
CMB temperature.  For higher $\Eg$, the integral drops sharply
as shown in fig.\ \ref{ICint}.  To get a contribution at the 
lowest Chandra energies $\Eg\cong 0.8$ keV, for example, would thus require 
$m_\chi \gtrsim 0.6$ GeV.

At energies below that of the CMB photons, 
${\overline E}_{\gamma'}/\z_d=2.7\, T_0$, 
(\ref{ICres}) scales like $\Eg^{-1/2}$ which translates
into a subdominant contribution of the form  $\Eg^{-3/2}$ 
to the temperature $T_\A$ of the excess radio background, where
\be
	T_\A = {\pi^2\over \Eg} {dn_\gamma\over d\Eg}
\ee
We can put an upper bound on the initial $\chi$ abundance such that
this extra contribution to the radio spectrum does not exceed
the observed $\Eg^{-3/2}$ contribution at the
highest energies where it is reliably observed.  Taking this to be
$10\,E_0 = 4.1\times 10^{-5}$ eV, corresponding to frequency $\nu = 10$
GHz, and the corresponding temperature $T_\A \cong 0.0032$K $= 2.8\times
10^{-7}$ eV (see fig.\ \ref{data}), we obtain the bound
\be
	Y_\chi < 2\times 10^{-7}\left(\z_d\over 1100\right)
\label{icradio}
\ee
on the relative initial abundance of $\chi$ to photons, $Y_\chi =
n_\chi/n_\gamma$, before the decays.
From (\ref{massbound}) one finds that for $m_\chi$ greater than a few
times $m_e$, the Compton contribution to the spectrum extends
well above the highest frequency $\sim 90$ GHz measured by
ARCADE-2 before cutting off.  For larger values of $m_\chi$ such
that the Compton contribution extends into the x-rays, we can
derive a much stronger constraint.

To compare (\ref{ICres}) with observations such as those of 
Chandra \cite{Hickox:2007gj}, we compute
\be
\int I_\nu d\nu = {1\over 4\pi}\int_{E_1}^{E_2}d\Eg \Eg {dn_{\gamma,0}\over d\Eg}
	\cong {85\, Y_\chi(\z_d/1100)^{-1} {\rm\ erg}
	\over{\rm cm}^{2}\,
	 {\rm s}\,{\rm deg}^2}
\label{Iobs}
\ee
in an energy window over which surface brightness has been measured
(multiplying by $3\times 10^{-4}$ 
to convert from str$^{-1}$ to deg$^{-2}$).
The value $85\, Y_\chi$ is for the $0.65-1$ keV energy interval,
in which the 
 measured value is $10^{-12}$ erg
cm$^{-2}$ s$^{-1}$ deg$^{-2}$.  We thereby derive a limit on the abundance
of decaying $\chi$ particles relative to photons of 
\be
	Y_\chi \lesssim 10^{-14}\left(\z_d\over 1100\right)
\label{Chandra}
\ee
which is much more stringent than (\ref{icradio}).  However
we emphasize that (\ref{Chandra}) applies only for 
$m_\chi \gtrsim 0.6$ GeV; otherwise Compton emission
in the $\Eg\sim$ keV region is kinematically blocked, since then
$E_0>E_d$ and the Compton spectrum cuts off as shown in fig.\
\ref{ICint}.   In this case, a stronger constraint might ostensibly
come from lower-energy x-ray observations, such as from XMM-Newton
\cite{xmm-newton}.  But in section \ref{data} we noted that photons
of energy lower than 1 keV are effectively attenuated through absorption
by hydrogen.  Hence the $m_\chi<$ 600 MeV restriction 
is sufficient to generally circumvent the constraints from
x-ray observations.

\section{Synchrotron emission}
\label{syncsect}

To derive the spectrum of synchrotron emission from the injected
electrons, we follow a similar procedure as for  Compton in the
previous section.  The rate of production of synchrotron photons from a single
electron of energy $E_e$ is given by \cite{Blumenthal:1970gc}
\bea
	{dN_\gamma\over dt d\Eg} &=& {1\over (4\pi)^2 \nu} 
	{d\Eg\over dt d\nu}\nonumber\\
	&=& {e^2 m_e^2\over \sqrt{3}\pi E_e^2}F(x)
\label{dNdtdE}
\eea
where $x = 2\Eg m_e^3/(3eB E_e^2)$ and 
$F(x)$ is the integral over 
harmonics of the gyrofrequency of the electron:
\bea
	F(x) &=& \int_x^\infty K_{3/2}(\xi) d\xi\nonumber\\
	&\cong& {4\pi\over 3^{1\over 2}2^{1\over 3}\Gamma({1\over
	3})}x^{-{2\over 3}}
	e^{-x}
\label{Fapprox}
\eea
Although $F(x)$ can be expressed exactly  in terms of hypergeometric
functions, we find the more tractable approximation in the second
line, which is just the known small-$x$ asymptotic value times
$e^{-x}$. This is accurate for $x\ll 1$  while overestimating the true
value by $\sim 20$\% near  $x\cong 2.5$ where $F$ is maximized, which
is sufficient for our purposes.  

\begin{figure}[t]
\centering
\vspace{-0.5cm}
\includegraphics[width=0.45\textwidth]{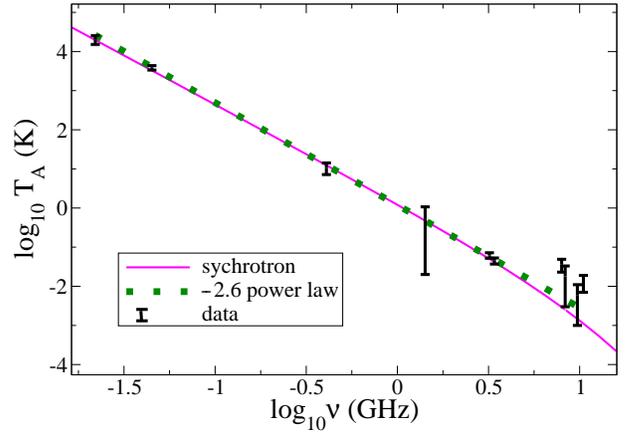}
\caption{Solid (magenta): the shape of the
predicted contribution to the radio background from 
synchrotron emission, when the constraint (\ref{mchiconst}) is saturated.}
\label{sync}
\end{figure}

The desired quantity ${dn_e/ dE_e}$ is obtained by multiplying
(\ref{dNdtdE}) by the electron density $n_e$ and integrating with
respect to time.  As in the previous section, the relevant time
dependence is that of the rapidly decreasing electron energy
(\ref{Eeeq}).  Again changing variables $t\to u=x^2$ and taking
$\Eg\to\z_d\Eg$ (note that also $B\to \z_d^2 B$ in order to reexpress
$B$ at $t_d$ in terms of its comoving value), we find
\bea
	{dn_{\gamma,0}\over d\Eg} &=&{0.048\,Y_\chi\,B^{3/2}\,m_e^{3/2}\over
	\sqrt{e}\z_d^{3/2}\,T_0\,\Eg^{3/2}} 
 \int_{\sqrt{C_\B}}^{\sqrt{C_\B}{E_d\over m_e}}\!\!\!\!\!\!du\, u^{2/3}
	e^{-u^2}
\label{sres}
\eea
where $C_\B = 2\Eg m_e^3/(3eB\z_d E_d^2)$.  We are interested in
parameter values such that the upper limit of the integral in 
(\ref{sres}) is $\gg 1$; in that case the integral can be approximated
by $\sfrac12$ the incomplete gamma function $\Gamma(\sfrac56,C_\B)$.
The latter approaches 1.13 as $C_\B\to 0$
but falls exponentially with large $C_\B$.  To insure that $T_\A\sim
\Eg^{-2.5}$, it is necessary to keep 
$C_\B\lesssim 1$
over the relevant range of $\Eg$.  This puts a constraint on the
injected energy, here from $m_\chi$,
\be
	m_\chi > 6.7{\rm\ GeV}\times \left({\z_d\over 1100}\right)^{-1/2}
	\left(B\over 50{\rm\ nG}\right)^{-1/2}
\label{mchiconst}
\ee
The distortion away from the pure $-2.5$ power law for the case where
$C_\B=1$, hence (\ref{mchiconst}) is saturated, is shown in fig.\ 
\ref{sync}.

Using $T_\A = (\pi^2/\Eg)\,dn_\gamma/d\Eg$ and normalizing to the
measured value at 45 MHz, we find that the required relative
abundance is
\bea
	Y_\chi \cong 6\times 10^{-12}
	\left({\z_d\over 1100}\right)^{3/2}
	\left(B\over 50{\rm\ nG}\right)^{-3/2}
\label{syncradio}
\eea
which is inconsistent with the Chandra constraint (\ref{Chandra})
unless $B$ exceeds the CMB bound.  To make them consistent,
we would need $B > 0.8\,\mu{\rm G}({\z_d/6})^{1/3}$.
Here we have expressed the result with lower redshifts in mind,
since this reduces the size of the required field somewhat.
This assumes that $m_\chi>600$ MeV so that (\ref{Chandra})
applies.  To evade the x-ray constraint by having $m_\chi<600$ MeV,
(\ref{mchiconst}) would require $B$ to be even larger,
$B > 6.2\,\mu{\rm G}(1100/\z_d)$.  However we shall see that
CMB constraints on charged particle injection require yet larger
values of $B$ than either of these bounds.

\section{Dark matter annihilations and nonsudden decays}
\label{suddensect}

In the previous sections, for simplicity we considered the case in
which electrons are injected suddenly at a specific time $t_d$
(the sudden decay approximation).  It is straightforward to generalize
this to a continuous injection of energy; one replaces $Y_\chi$ by
\be
	Y_\chi \to \int dt_d {dY_\chi\over dt}
\ee
(taking into account the time-dependence of $\z_d$ in the expressions
for $dn_\gamma/d\Eg$)
where $dY_\chi/dt$ is the relative abundance per unit time of
$\chi$ particles that decay into electrons.  The same logic can be
applied to annihilations $\chi\chi\to e^+ e^-$.  The expressions for
$dY_\chi/dt$ for decays or annihilations are, respectively,
\be
	{dY_\chi\over dt} = \left\{\begin{array}{ll}
	Y_{\chi,0}\,\tau^{-1}e^{-t/\tau}, & {\rm decay}\\
	& \\
	Y_{\chi,0}^2\, n_\gamma\langle\sigma v\rangle,
	& {\rm annihilation}\end{array}\right.
\ee

Let us consider the synchrotron photons coming from annihilation 
$\chi\chi\to e^+e^-$.  We can convert the integral over $t_d$ into one over $z_d$
using $t_d\cong t_0 z_d^{-3/2}$, which is not accurate at late times
when dark energy dominates, but is adequate during matter domination.
Carrying out the integral over $z_d$, where for a first estimate
we approximate
$\Gamma(5/6,C_\B)$ as a step function that goes to zero at 
$C_\B=1$, we find that $dn_{\gamma,0}/d\Eg$ in
(\ref{sres}) is modified by replacing
\be
	{Y_\chi\z_d^{-3/2}} \to \sfrac32 Y_{\rm \chi,0}^2\, t_0\, n_{\gamma,0}
	\langle\sigma v \rangle 
	\ln\left(\z_{\rm dec}\over\z_{\rm end}\right)
\label{annY}
\ee
where $\z_{\rm end}$ is the value of $\z_d$ such that $C_\B=1$ and
$\z_{\rm dec}=1100$.
Here $Y_{\chi,0}$ is the dark
matter abundance after it has frozen out, which we can rewrite as
$Y_{\chi,0} = n_{\chi,0}/n_{\gamma,0} = 2.7 T_0\Omega_\chi/
(m_\chi\Omega_\gamma)$, and $\langle\sigma v\rangle$ is the cross
section to produce $e^+e^-$.  It is therefore no more than 
$(\Omega_{\sss DM}/\Omega_\chi)\langle\sigma
v\rangle_0$, where $\langle\sigma
v\rangle_0 = 3\times 10^{-26}$ cm$^3$/s is the standard thermal cross
section for obtaining the observed $\Omega_{\sss DM} =
0.22$.\footnote{If there is some other annihilation channel
determining the $\chi$ relic density, then $\langle\sigma
v\rangle$ for $\chi\to e^+e^-$ can be smaller than this bound.} Here
we only assume that $\Omega_\chi < 0.22$ since $\chi$ need not be the
only species of dark matter.  Now equating the modified (\ref{sres}) 
to the observed radio spectrum, we find the necessary 
condition
(but not sufficient, as will become immediately apparent)
\be
	m_\chi = 0.6 {\rm\ MeV} \times \sqrt{r}
	\left(\Omega_\chi\over\Omega_{\sss DM}\right)^{1/2}
	\left(B\over 50 {\rm\, nG}\right)^{3/4}
\label{annlimit}
\ee
where we define $r$ to be the ratio of $\langle\sigma v\rangle$
for $\chi\chi\to e^+e^-$ to the total annihilation cross
section.  For simplicity we assume that any additional annihilation
channels do not result in charged particles, since otherwise these
will quickly decay into electrons and behave similarly to the
electrons produced as primary annihilation products.  The
requirement (\ref{annlimit}) insures that the spectrum matches the
observed one at arbitrarily low frequencies, but it does not take
into account the need for the high-frequency cutoff to extend to the
observed GHz region.  This requires 
$\ln(\z_{\rm dec}/\z_{\rm end})\sim 1$, which is only consistent if
$\z_{\rm end}\lesssim \z_{\rm dec}$, which puts a second constraint
on the mass (see (\ref{mchiconst})),
\be
	m_\chi>6.7{\rm\, GeV}\times 
	\left(B\over 50 {\rm\, nG}\right)^{-1/2} 
\ee
These constraints begin to overlap (leaving no allowed parameter
space) if the field is too small, leading
to the lower bound
\be
	B > 90\,\mu{\rm G} \times
	\left(\Omega_{\sss DM}\over r\Omega_\chi\right)^{2/5}
\label{annBlimit}
\ee 
In the case where this bound is saturated, 
the mass is $m_\chi=160\,(r\Omega_\chi/\Omega_{\sss DM})^{1/5}$ MeV.  We will see that
this is strongly ruled out by CMB constraints on charged particle 
injection.  Thus dark matter annihilations in the early universe
cannot explain the radio excess even with late-time magnetogenesis.

For decays, we can carry out the time-averaging to find the more
accurate dependence of the normalization of the spectra 
upon $\z_d$.  This results in a correction factor of order unity:
\be
	{1\over \z_d^p}\to {1\over \z_d^p}
	\int_{t_{\rm dec}/\tau}^{t_0/\tau} dx\, x^{2p/3} e^{-x}
\ee
where $t_{\rm dec} = 3.8\times 10^5$ y is the age of the universe at
decoupling, and $t_0=13.8$ Gyr is the present age.  For $t_{\rm dec}
\ll \tau \ll t_0$, the correction factor is approximately $\Gamma(1 +
2p/3) = 0.9,\,1$ for $p=1,\,3/2$, corresponding to the cases of
Compton and synchrotron emission, respectively.  Thus the sudden approximation works
quite well for the case of decays when $\tau$ is in the range of
interest.

\section{CMB constraints on charged particle injection}
\label{cmbsect}

\begin{figure}[t]
\includegraphics[width=0.51\textwidth]{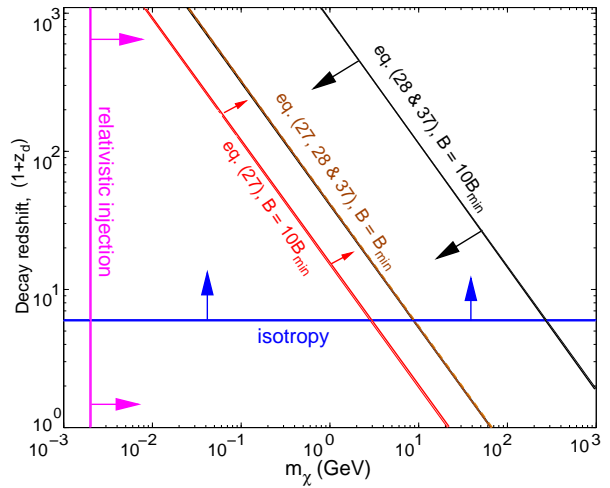}
\caption{Summary of constraints on models with electron
injection at redshift $z_d$ from $\chi\to e^+e^-$ decays, as a
function of $m_\chi$.   Allowed regions are in directions of arrows.
``Isotropy'' (blue) denotes the limit $z_d>5$ due to the isotropy of
the signal. ``Relativistic injection'' (magenta) denotes requirement 
limit $m_\chi\gtrsim 2$ MeV in order to obtain relativistic electrons.
The ``eq.\ (27, 28 \& 37)'' line (brown)  is the coincidence of upper and
lower limits on $m_\chi$ in the case of minimum $B$ field given
by eq.\ (\ref{finalBbound}). 
If $B$ is allowed to be 10 times larger than this minimum value, 
the allowed region is bounded by the 
``eq.\ (27)'' (red) and ``eq.\ (28 \& 37)'' (black) curves, coming from
the shape of the synchrotron spectrum and the CMB limit on charged
particle injection, respectively.}
\label{constraints}
\end{figure}

There are stringent bounds on the injection of ionizing energy after
decoupling due to its distortion of the CMB anisotropies, either
through particle decays \cite{Chen:2003gz,Zhang:2007zzh} or
annihilations \cite{Galli:2009zc,Slatyer:2009yq}.  These depend 
mainly upon the total amount of kinetic energy of charged particles 
that is injected, and so are closely correlated with the requirements
for producing the observed radio background.   
We first consider the case of decaying particles.  Ref.\ \cite{Zhang:2007zzh} 
obtains 
\be
	{Y_\chi\over\tau} \lesssim 2\times 10^{-25} {\rm s}^{-1}
	 Y_{\sss DM}
\ee
for $\tau > 10^{17}$ s, and a somewhat less stringent limit at shorter
lifetimes; however the above approximation will be adequate for
our purposes since the tension of this constraint with what is needed
for the radio excess is minimized at the higher values of $\tau$, 
despite the modest weakening of the constraint at low $\tau$. 
Here $Y_{\sss DM} = (\Omega_{\sss DM}/\Omega_{\gamma'})(
\overline{E}_{\gamma'}/ m_\chi)$ is the standard abundance that $\chi$ would have if
it was a stable particle constituting the full WMAP DM abundance. 
Substituting $\z_d$ for $\tau$ we then find
\be
	Y_\chi< 2\times 10^{-17}\left(6\over\z_d\right)^{3/2}
	\left({1{\rm\ GeV}\over m_\chi}\right)
\label{CMB_bound}
\ee
We have taken $\z_d=6$ to be the latest decay epoch consistent with
generating the signal before significant clustering of the DM would
occur and induce unwanted fluctuations.  If we take the synchrotron
mass bound (\ref{mchiconst}) to be saturated, then (\ref{CMB_bound})
becomes $Y_\chi < 10^{-18}(6/\z_d)^{1/2}(B/1\mu$G$)^{1/2}$.  Comparing
this to the required value of $Y_\chi$ (\ref{syncradio}), we get the
lower bound on $B$
\be
	B > 5\,\mu{\rm G}\times \left(\z_d\over 6\right)^{5/4}
\label{finalBbound}
\ee
For this value of $B$, the minimum $\chi$ mass from (\ref{mchiconst})
is $m_\chi = 9\times(\z_d/6)^{-9/8}$ GeV, and the abundance is $Y_\chi = 
2\times 10^{-18}(\z_d/6)^{-3/8}$.  This is compatible with the x-ray
constraint (\ref{Chandra}), which can be written as $Y_\chi < 5\times
10^{-17}(\z_d/6)$. A summary of the bounds on electron injection from $\chi$ decay is presented in Figure \ref{constraints}.

For $\chi\chi\to e^+e^-$ annihilations, we adapt the result of 
ref.\ \cite{Slatyer:2009yq} to obtain 
\be
	\langle\sigma v\rangle
\equiv r\langle\sigma v\rangle_{\rm tot} \lesssim f^{-1}\langle\sigma v\rangle_0
	\left(m_\chi\over 10{\rm\, GeV}\right)\left(\Omega_{\sss
	DM}\over\Omega_{\chi}\right)^2
\label{slatyer}
\ee
where $f$ is an efficiency factor that is close to 1 for annihilation
into electrons, and $r$ was defined below eq.\ (\ref{annlimit}).  
Since the rate of injection of charged particles depends upon
$n_\chi^2$, we have appended the factor 
$(\Omega_{\sss
DM}/\Omega_{\chi})^2$ for the case $\Omega_{\chi} < \Omega_{\sss DM}$
to account for the weakening of the bound derived by 
\cite{Slatyer:2009yq}, where it was assumed that 
$\Omega_{\chi} = \Omega_{\sss DM}$.
Consider the case where the $B$ field is just barely large enough
to give the radio excess, so that (\ref{annBlimit}) is saturated.
Then using the fact that $\langle\sigma v\rangle_{\rm tot}/
\langle\sigma v\rangle_{0} = \Omega_{\sss DM}/\Omega_\chi$ and
the value of $m_\chi$ determined below eq.\ (\ref{annBlimit}), we find that the CMB
constraint can be written as
\be
	\left(r\Omega_\chi\over \Omega_{\sss DM}\right) <
	6\times 10^{-3}
\ee
which with eq.\ (\ref{annBlimit}) then implies $B > 3$ mG (milligauss)
in order for annihilations to give the radio excess.  This is such a
strong field that it would overclose the universe.

\section{Millicharged model}
\label{model}

If we insist that $B$ was generated prior to decoupling and thus must 
obey the CMB constraints, additional new physics is required. In this
section we present a somewhat more complicated model, in which the
injected particles, instead of being electrons, are exotic particles
that we denote by $\tilde e$, carrying a fractional electric charge
$\epsilon e$. In adddition, we assume that $\tilde e$ has a larger
charge $g$ under a hidden, unbroken $U(1)'$ gauge group that comes
with its own ``dark photons,'' (denoted by $\tilde\gamma$) having a
cosmological background analogous to the CMB. This scenario has been
considered previously in the millicharged atomic dark matter model of
ref.\ \cite{Cline:2012is,Cline:2012ei}.

It is straightforward to generalize the  Compton result (\ref{ICres})
to the present model.  For the rate of production of CS photons, 
the combination $e^4 n_\gamma {\overline E}_{\gamma'}^{1/2}$ coming from
$e^4 n_\gamma/( {\overline E}_{\gamma'} E_0^3)$ is replaced by 
\be
	A\equiv
	g^2 (\epsilon e)^2
	{n_{\tilde\gamma}\overline{E}_{\tilde\gamma}^{1/2}}
	+ (\epsilon e)^4 {n_\gamma{\overline E}_{\gamma'}^{1/2}}
\label{eq1}
\ee
On the other hand in the energy loss rate that eventually appears in
the denominator, the combination $e^4 \rho_\gamma$ is replaced by 
\be
	A' = g^4 \rho_{\tilde\gamma} + g^2 (\epsilon e)^2(\rho_{\tilde\gamma}
	+ \rho_\gamma) + (\epsilon e)^4 \rho_\gamma
\label{eq2}
\ee
Thus the CS spectrum will be proportional to $A/A'$.  In addition 
the electron mass is replaced by $m_{\tilde e}$. Assuming that the
first term in each of (\ref{eq1}) and (\ref{eq2}) dominates,
we find that the x-ray constraint (\ref{Chandra}) generalizes to 
\be
	Y_\chi < 5\times 10^{-17}\left(\z_d\over 6\right)
	\left({\tilde T_0\over T_0}\right)^{1/2}\left(g\over\epsilon
e\right)^2 \left(m_e\over m_{\tilde e}\right)
\label{icradio2}
\ee
In contrast, the CMB constraint depends mainly upon the fraction of
ionizing electromagnetic energy per decay.  In the present model,
this fraction is given by $(A'-g^4 \rho_{\tilde\gamma})/A'$.
The bound (\ref{CMB_bound}) is accordingly weakened to become
\be
	Y_\chi< 2\times 10^{-17}\left(6\over\z_d\right)^{3/2}
	\left({1{\rm\ GeV}\over m_\chi}\right)\left(g\over\epsilon
	e\right)^2\left({\tilde T_0^4 \over  T_0^4}\right)
\label{CMB_bound2}
\ee
where we have assumed that the leading terms in numerator and
denominator of the reduction factor are dominant, and also that
$\tilde T_0^4\ll  T_0^4$.  These two assumptions would be
consistent if for example $\tilde T_0^4\sim 0.1 T_0^4$, the maximum
allowed by constraints on the Hubble rate during big bang
nucleosynthesis.  We need to choose the more restrictive of the 
two bounds (\ref{icradio2}) and
(\ref{CMB_bound2}) in what follows.

The millicharged model can also admit a primordial dark magnetic
field $\tilde B$ whose magnitude is less constrained than that of the
visible $B$ field, allowing for the possibility of early $\tilde B$
magnetogenesis that is unconstrained by the CMB.  The modified synchrotron spectrum can be deduced from 
the derivation of section \ref{syncsect} by making the following
changes.  (1) $e\to \epsilon e$ for the rate of emission in 
(\ref{dNdtdE}); (2) $eB\to g\tilde B$ in the definition of $x$, since
the cyclotron frequency is now determined by this combination;
(3) $e\to g$ in the Thomson cross section 
and $\rho_{\gamma}\to \rho_{\tilde\gamma}$ coming from the energy 
loss rate, which is dominated by $\tilde\gamma$ Compton emission;
(4) obviously $m_e\to m_{\tilde e}$ everywhere as well.
In this way, we
find that the  required abundance (\ref{syncradio}) generalizes to
\bea
\label{syncrat}
	Y_{\chi,\rm sync} &=& 2.7\times 10^{-17}\times\\
	&&\epsilon^{-2}\left({\tilde T_0\over T_0}\right)^4\!\!\!
	\left(\z_d\over 6\right)^{\!\!\frac32}\!\left(\tilde B\over 
	1\,\mu{\rm G}\right)^{\!\!-{3\over 2}}\!\!\!\left(m_e\over m_{\tilde
e}\right)^{\!\!{3\over 2}}
	\left(g\over e\right)^{\!\!{5\over 2}}\nonumber
\eea
which is valid as long as $g^4\,\tilde T_0^4\gtrsim (g\epsilon e)^2\, T_0^4$.
Similarly, the generalization of the mass constraint 
(\ref{mchiconst}) is
\be
	m_\chi \gtrsim 20\,{\rm GeV}\left(\z_d\over 6\right)^{-\frac12}
\left(\tilde B\over 1\,\mu{\rm G}
	\right)^{\!\!-\frac12}\left(e\over g\right)^{\!\!\frac12}
\left(m_{\tilde e}\over m_{e}\right)^{\!\!\frac32}
\label{mchiconst2}
\ee

Proceeding as in section \ref{cmbsect}, by inserting
(\ref{mchiconst2}) into (\ref{CMB_bound2}) and comparing to 
(\ref{syncrat}), we find that all the dependence upon new parameters
cancels out and yields (\ref{finalBbound}) again, but now as a
constraint on $\tilde B$ instead of $B$.  The advantage is that there
are relatively few direct constraints on the dark $\tilde B$ field, and
it could therefore have been generated prior to decoupling.  
For example taking $\z_d=6$, 
$m_\chi=5\times(m_{\tilde e}/m_e)^{3/2}$ GeV,
$\tilde B = 5\mu$G, $\epsilon = 10^{-3}$,
$g = \sqrt{10}e$,  we find from 
(\ref{syncrat}) that $Y_\chi = 4\times 10^{-12}
(m_e/m_{\tilde e})^{3/2}$, which marginally satisfies the CMB
constraint (\ref{CMB_bound2}) for any value of $m_{\tilde e}$.
We can then take $m_{\tilde e} > 100$ MeV (hence $m_\chi>14$ TeV) and be consistent with
constraints on millicharged particles 
\cite{Davidson:2000hf,Jaeckel:2010ni,Prinz:1998ua}.  The x-ray
constraint (\ref{icradio2}) is weaker than (\ref{CMB_bound2})
and provides no additional information.  For such values of
$Y_\chi$,  the mass density of $\chi$ is sufficiently small before
it decays, $\Omega_\chi \sim 0.04\,\Omega_b$, that there is no danger
of it significantly affecting the expansion history of the universe. 

\section{Conclusions} 
\label{conc}

The diffuse excess radio background observed by the ARCADE-2
collaboration has a spectrum that appears to be consistent with
observations of four other groups.   This adds to the plausibility of
it being a real anomaly, and the observation of  ref.\
\cite{Holder:2012nm} concerning its homogeneity seems to constitutes a
strong motivation for its origin to be cosmological in nature, in the
absence of ubiquitous astrophysical sources of size $> 1$ Mpc.

Using approximate analytic methods,\footnote{notably the approximation
of (\ref{Fapprox}) and extending the upper limit of integration
in (\ref{sres}). We expect this to introduce errors of order
20\% in the normalization of the spectrum, and O(1) errors in the
details of the spectrum in the high-frequency region where it is 
rapidly falling away from the $-2.5$ power law behavior.}
we have shown that injection of relativistic electrons after
decoupling, through late decays of a metastable particle $\chi$,  can
give rise to a radio excess with the correct power law fall-off
through synchrotron emission.  However we find that injecting enough
charge to get the right normalization of  the excess background 
should also have distorted the CMB fluctuations relative to their
observed properties, due to the deposition of electromagnetic energy
into the intergalactic medium.  This contradiction can be circumvented
if the primordial $B$ field is larger than expected on the basis of
other CMB constraints, $B > 5\,\mu$G.  The size of the required field
is smallest when the decays are as late as possible (but still before the
onset of structure formation, to avoid large fluctuations in the radio
background).  Such large fields would have to be generated by some
unknown cosmological mechanism after decoupling in order to avoid
the CMB $B$-field constraint, $B\lesssim$ several nG.

Based on our results, the simplest and most conservative
cosmological scenario  for producing the excess radio background is
through the synchroton radiation emitted by electrons from 
decays $\chi\to e^+e^-$ at redshift $z=5$, in a magnetic field
of strength $B\sim 5\mu$G, with $m_\chi \gtrsim 9$ GeV and an initial
abundance $Y_\chi \sim 10^{-18}$ relative to photons. Earlier
decay epochs would require larger $B$ fields, given by
(\ref{finalBbound}).   This requires magnetogenesis at low redshifts
$z<1100$ to circumvent CMB constraints on the primordial magnetic
field.  We find that the annihilation scenario, $\chi\chi\to e^+e^-$,
is too inefficient and would require unacceptably large $B$ fields to
produce the observed radio excess. 

One might wonder whether decays into other known charged particles
than the electron could give more promising results, since the CMB
bound is weakened in that case. The weakening comes about because some
of the initial energy eventually goes into neutrinos, which do not
ionize the intergalactic medium. We do not expect any improvement in
this way however, since the same loss of efficiency in the production
of radio and other electromagnetic backgrounds should occur, and the
amount of radiation produced by the unstable charged particles before
they turn into electrons must be very small due to their short
lifetimes, compared to the time scale for energy loss through CS.

To demonstrate an alternative example that does not require late-time
magnetogenesis, we presented a model of small fractionally charged
($\sim 10^{-3}e$) decay products with mass $\gtrsim 100$ MeV, produced
from the decays of heavy $\gtrsim 14$ TeV parent particles.  If the
fractionally charged particles undergo cyclotron motion due to their
interaction with a dark background magnetic field $\tilde B \gtrsim
5\mu$G, the resulting synchrotron emission (in normal photons) can
explain the radio anomaly without conflicting with CMB observations. 
The fractionally charged particles are stable relics, but with an
abundance less than $10^{-5}$ that of baryons, making them innocuous
at present times. This example might be regarded as proof that it is
possible to  construct a viable, albeit exotic model; we hope it will
motivate others to search for alternative
possibilities.\footnote{Recently ref.\ \cite{Lawson:2012zu} appeared,
which also proposes a cosmological dark matter mechanism.}

\smallskip

{\bf Acknowledgments.} We are grateful to F.\ Tavecchio for pointing
out an important error in the first version of this manuscript.
 We thank M.\ Cirelli, R.\ Durrer, F.\ Finelli, M.\ Giovannini,
 G.\ Holder, Z.\ Liu, T.\ Kahniashvili, M.\ Kamionkowski, 
G.\ Moore, V.\ Poitras, P.\ Scott and T.\ Slatyer for helpful dicussions or correspondence.
Our work was supported by the Natural Sciences and Engineering Research Council
(NSERC, Canada).   AV is supported by European contract
PITN-GA-2011-289442-INVISIBLES and 
Fonds de recherche du Qu\'ebec - Nature et technologies
(FQRNT).

\appendix
\section{Stochastic $B$ field from $\chi\to e^+e^-$ decay}
\label{Bdecay}
In this appendix we estimate the stochastic magnetic field spectrum that is
generated from the decays of a heavy particle $\chi$ into $e^+e^-$.
Consider a decay at time $t=0$ and position $\vec r=0$, producing an
electron of velocity $\vec v$.  The vector potential due to the $e^+e^-$ pair
is
\be
	\vec A = e\vec v\left(|\vec r-\vec v t|^{-1} - |\vec r+\vec v
t|^{-1}\right)
\ee 
We can compute the desired correlation function $\langle B_i(x) B_j(y)\rangle$
from $\langle A_i(\vec x) A_j(\vec y)\rangle$ by taking the appropriate curls.  The
correlators can be found by averaging over a random ensemble of vector
potentials due to currents
emanating from arbitrary points in arbitrary directions (but for simplicity we
take them all to arise at the same time).  We take the contributions from
different currents to be uncorrelated so that only diagonal terms in the
product of the two $A$'s contribute.   The correlator thus becomes a sum over
all particles, averaging over the position of the decaying parent and the
direction of the products.  Let $N$ be the total number of decaying particles
in a volume $V$, and define $\vec x_\pm = \vec x - (\vec x_0\pm\vec v t)$ where $x_0$ is the initial position
of a fiducial pair; similarly for $\vec y_\pm$.  Then
\bea
	\langle A_i(\vec x) A_j(\vec y)\left\rangle = e^2 
N\Big\langle v_i v_j\right.
	&\!\!\!\!\Big(&
	\left(|\vec x_+|^{-1}- |\vec x_-|^{-1}\right)\quad\quad\quad\\
&\times& \left(|\vec y_+|^{-1}- |\vec y_-|^{-1}\right)\Big)\Big\rangle\nonumber
\eea
The averaging means
\be
	\langle X \rangle = {1\over 4\pi V}\int d^3 x_0\int d\Omega_v X
\ee
The integral over $x_0$ can be done by expressing
\be
	|\vec x_\pm|^{-1} = {1\over\sqrt{\pi}}\int_0^\infty 
	{ds_\pm\over\sqrt{s_\pm}}
	e^{-s_\pm|\vec x_\pm|^2}
\ee
The integrals over $s_\pm$ can also be done exactly, leading to
\be
	\langle A_i(\vec x) A_j(\vec y)\rangle = e^2 n_e\int d\Omega_v 
	v_i v_j \left(|\vec x-\vec y|-|\vec x_+-\vec y_-|\right)
\ee
where $n_e$ is the electron density.
The remaining integral over angles can also be done exactly, but at large times
the leading contribution to the gradients will be from the first term, which is
given by 
\be
	\langle A_i(\vec x) A_j(\vec y)\rangle \cong 4\pi e^2 v^2 n_e
	\delta_{ij}|\vec x - \vec y|
\ee
The $B$ field correlator thus goes like $1/|\vec x-\vec y|$, corresponding to a
spectral index of $n_\B = -2$ in the notation of section \ref{inject}.

The smallest scale that can be relevant for synchrotron emission
is the Larmor radius $r=m_e c/eB$.  Eliminating $r$ from this equation
and (\ref{decayB}) at $r=|x-y|$, we find the field strength
$B = 4\pi  e^3 n_e/m_e$ (evaluated at the decay redshift 
$z_d < 1100$; this value of $B$ must be divided by 
$z_d^2$ to convert it to the comoving value for comparison with the
CMB constraint).  Even if the electrons are injected
with unit abundance relative to photons, this is orders of
magnitude smaller than the CMB limit and thus irrelevant. 

\section{Brehmsstrahlung}
\label{brem}
The final mechanism for photon production from charged particles is by
Bremsstrahlung with the ambient neutral (before reionization) or charged (after)
hydrogen atoms. We obtain the photon spectrum in the same way as before, by
assuming a sudden decay, and replacing the integral over time with an integral
over the electron's energy, which is lost due to Compton scattering. The bremsstrahlung spectrum of a relativistic electron colliding with a hydrogen atom is 
\cite{Blumenthal:1970gc}:
\bea
  \frac{\ud N_\gamma}{\ud E_\gamma dt}  
&=&\frac{\alpha r_0^2}{E_\gamma E_e^2}
\left[(E_e^2 + (E_e-E_\gamma)^2)\phi_1\right.\nonumber\\
 &-&\left. \frac{2}{3} E_e (E_e-E_\gamma) \phi_2 \right],
\eea
where $r_0$ is the classical electron radius and $E_e - E_\gamma$ is approximately the electron energy after scattering. The functions $\phi_i$ parametrize the atomic shielding. When $E_\gamma \ll E_e$, there are two possibilities: 1) scattering off unshielded charges (\textit{e.g.} in an ionized universe):
\begin{equation}
 \phi_1 = \phi_2 = 4 \ln \left(\frac{E_e(E_e-E_\gamma)}{E_\gamma m_e} \right) -2;
\end{equation} 
or 2) neutral hydrogen, in which case the shielding goes to:
\begin{equation}
 \phi_1 \simeq \phi_2 \simeq 45.
\end{equation} 
Then, with $E_e \gg E_\gamma$ we get the result for unshielded charges:
\bea
 \frac{\ud n_\gamma}{\ud E_\gamma} &=& 
\frac{2 n_\chi n_H \alpha r_0^2 m_\chi }{3 C_e E_\gamma}\frac{4}{3} 
\Bigg(\frac{6}{m_e} + \frac{4}{m_e}\ln\left(\frac{m_e}{E_\gamma}
\right)\nonumber\\
 &-&  \frac{8}{m_\chi} \ln \left( \frac{m_\chi^2}{4E_\gamma m_e}
\right) \Bigg)
\eea 
and for shielded charges:
\begin{equation}
 \frac{\ud n_\gamma}{\ud E_\gamma} = \frac{2 n_\chi n_H \alpha r_0^2 m_\chi}{3 C_e m_e E_\gamma}\frac{4}{3}
\end{equation} 
We have taken the integration limits from $m_\chi/2$ to $m_e$ (strictly speaking the lower limit should be $m_e + E_\gamma$, but the latter correction is negligible at small $E_\gamma$). Note that in both cases the spectrum goes as $E_\gamma^{-1}$, which does not match the desired shape to reproduce the radio excess. We finally estimate the contribution to other limits. Rewriting the latter in the language of Section 5, and evaluating the spectrum today:
\begin{equation}
\frac{\ud n_\gamma }{\ud E_\gamma} = \frac{3 \alpha n_{\chi,0} m_e}{4
\pi m_p z_d E_\gamma }\frac{\Omega_{B,0}}{\Omega_{\gamma,0}}
\end{equation}
Comparing this with the CS, this is a very small contribution, even in the keV region:
\begin{equation}
\frac{I_{\rm{Brems}}}{I_{\rm{C}}} \simeq 4 \times 10^{-10} \left(\frac{E_\gamma}{\rm{eV}} \right)^{1/2}.
\end{equation}
If injection of electrons instead occurs after reionization, the bremsstrahlung contribution is at most larger by a factor of $2(3+2 \ln (m_e/z_dE_\gamma)) \sim 80$ for $z_d \sim 10$ and $\nu \sim $ GHz, but is further suppressed at high energies due to the logarithmic terms.


\bibliographystyle{apsrev}

\end{document}